\documentclass{osa-article}
\usepackage{soul}
\usepackage{float}

\journal{osajournal}
\articletype{Research Article}
\begin{document}

\title{Mechanisms of the Nonlinear Refractive Index in Organic Cavity Polaritons}

\author{Samuel Schwab, \authormark{1} William Christopherson, \authormark{1} Michael Crescimanno, \authormark{2} and Kenneth Singer\authormark{1,*}}

\address{\authormark{1}Physics Department, Case Western Reserve University, 10900 Euclid Ave, Cleveland, OH 44106\\
\authormark{2}Physics Department, Youngstown State University, 1 University Plaza, Youngstown, OH 44555}

\email{\authormark{*}corresponding email} %% email address is required


%%%%%%%%%%%%%%%%%%% abstract 

\begin{abstract}
 The nonlinear optical response of organic polaritonic matter has received increasing attention due to their enhanced and controllable nonlinear response and their potential for novel optical devices such as compact photon sources and optical and quantum information devices.  Using  z-scans at different wavelengths and incident powers we have studied the nonlinear optical dispersion of ultrastrongly coupled organic cavity polaritons near the lower polariton band.  We show that the up to 150-fold enhancement of the nonlinear response compared to a cavity-less organic film arises from an intensity-dependent polaritonic resonant frequency shift ("blueshift"). Consequently, we find that these z-scan data can only be described by several terms of a power series expansion in intensity whose respective contributions depend on power broadening and detuning from the lower polariton band.  We further show that the nonlinear response can be quantitatively described by a three-level molecular quantum model coupled to the cavity in which saturation reduces the Rabi splitting, thus accounting for the lower polariton band's observed blueshift.
\end{abstract}

%%%%%%%%%%%%%%%%%%%%%%%%%%  body  %%%%%%%%%%%%%%%%%%%%%%%%%%
\section{Introduction}
Cavity polaritons are light-matter mixed states arising from coupling between excitonic matter and confined light fields \cite{Sanvitto2016}. Being part exciton and part photon, polaritons are highly tunable and have thus inspired an emerging field for potential chemical and quantum engineering applications \cite{Schneider2017,Carusotto2013,Dovzhenko2018}. Already, a plethora of classical and quantum phenomena ranging from angle-dependent amplification \cite{Savvidis2000}, parametric oscillation \cite{Ferrier2010}, and enhanced emission \cite{Ballarini2014, Berghuis2019} to single-quanta entanglement preservation \cite{Cuevas2018}, room-temperature out-of-equilibrium Bose-Einstein condensation \cite{Daskalakis2014}, and superfluidity \cite{Lerario2017} have been found in polaritonic systems. Moreover, the ultrastrong coupling regime (where coupling energy compares to bandgap of material) opens up new avenues in tunability suitable for quantum and classical applications beyond the rotating-wave approximation \cite{FriskKockum2019,Forn-Diaz2019}. 

Cavity polaritons made from Wannier-Mott excitons found in semiconductor structures have  been studied extensively and typically fall within the strong coupling regime for light-matter interactions \cite{Luk2007}. In organic materials, the large binding energies of Frenkel excitons make ultrastrongly coupled polaritons easily attainable even with low-Q mirrors. Further, the ease of adjusting the exciton-photon coupling through mixing of organic dyes with polymers adds an additional and useful tunability \cite{FriskKockum2019, Liu2015, Anappara2009, Gambino2014}.

Exploration into the nonlinear optical properties of cavity polaritons has been fruitful. So far, enhanced and tunable third-harmonic generation \cite{Barachati2018, Liu2019, Wang2020}, enhanced second harmonic generation \cite{Chervy2016}, and novel four-wave mixing processes \cite{Romanelli2007} have been explored. These works suggest that the nonlinear response is dictated by the polariton states, instead of the intuitive material resonant states. In this work, we further elaborate on this picture for the sum-over-states model by investigating ultrastrongly coupled organic polariton states extensively using a z-scan technique over ranges of pump wavelengths and powers. We measure a nearly 150-fold enhancement of the nonlinear response around the lower polariton (LP) resonance when compared to the bare excitonic film. We quantitatively show that optical  saturation of the exciton reservoir leading to a blueshift of the LP band explains much of the experimentally observed behavior. A two-level Lorentzian model of the LP resonance indicates expected contributions of higher order nonlinear response to the intensity dependent refractive index. We find that a three-level quantum optics model of the organic molecule coupled to a cavity mode is able to quantitatively explain our observations. Notably, doing so with a three-level model only meaningfully introduces a single free parameter, an excited state mixing rate. 

\section{Background}

\noindent Recently, polaritonic systems containing Frenkel excitons have been shown to exhibit strong blueshift in the LP when pumped around the threshold for creating a polariton condensate \cite{Yagafarov2020}. This shift has been primarily attributed to saturation and intermolecular energy migration. Although our experiments are at excitations below this condensate regime, we explore the effects of saturation on the detectable output of our samples in a z-scan configuration and bring insight into how saturation plays a pivotal role in this nonlinear optical regime. 

As a simple model of how a resonant energy shift results in an intensity dependent refractive index, consider a two-level Lorentzian model comprised of a ground and excited state, in this case, the LP state. Denote the associated population loss and dephasing rates as $\gamma $ and $\gamma_2$, and the product of the dipole matrix element connecting these two states and the exciting electric field by   $E(t) = E e^{-i\omega t} + c.c.$, with $\omega$ representing the input frequency. Solving the optical Bloch equations in the rotating wave approximation for the steady state we arrive at the usual solution for the density matrix element associated with a transition between ground and excited state, $\rho_{eg}$. In the low intensity, low density limit, the real part of this solution gives the index of refraction as, ($I=E^2$)

\begin{equation}
    Re(\rho_{eg})-1=n-1=\frac{\omega N d^2 \delta}{2 \epsilon_0 c k_0 (\delta^2 + \gamma_2^2 + 4\frac{\gamma_2}{\gamma}E^2)}\Rightarrow U(G,\omega)\frac{\delta+\tau I}{(\delta+\tau I)^2+\Gamma^2 +PI},
\label{twoLevel_n}
\end{equation}

\noindent where $N$ represents the number density of the molecules, $d$ their dipole matrix element, $c$ the speed of light, $k_0$ is the vacuum wave vector, $\gamma$ is the population relaxation rate, $\gamma_2$ is the decoherence rate, and $\delta = \omega - \omega_0$ is the detuning from resonance. We simplify the expression using $U(\omega)$ a prefactor depending on the oscillator strength, and anticipating its significance for this study, we introduce an intensity-dependent blueshift parameter, $\tau$, into the detuning, $\delta \rightarrow \delta + \tau I$.  We also introduce $\Gamma$ as a composite loss parameter and $P$ as a power broadening parameter. Subsequently  expanding Eq. \ref{twoLevel_n} in powers of $I$  reveals various orders of nonlinear optical response.  Defining $\Delta n= n(I)-n(0)$ ,

\begin{equation}
\frac{\Delta n}{I}=n_2+n_4 I + n_6 I^2
\label{dn_overI}
\end{equation}

\noindent where

\begin{equation}
    n_2=U(\omega)\frac{(\Gamma^2 -\delta^2 )\tau-\delta P}{(\delta^2 + \Gamma^2)^2}
    \label{Deltan}
\end{equation}
Expressions for $n_4$ and $n_6$ are given in supplemental information. These expressions indicate how the various nonlinear orders depend on the excitation wavelength.  Note in the limit of small blueshift  ($\tau \approx 0 $) the expansion of the nonlinear index leads to sign alternation between nonlinear indices of increasing power as $n_2<0, n_4>0, n_6<0$. 
\noindent whereas at negative detuning for $\tau=0$ the alternation is reverse, that is $n_2>0, n_4<0, n_6>0$.
In this limit, this two-level model also indicates a general qualitative trend;  that with blueshift parameter ($\tau$)  each nonlinear term is not necessarily an odd function of the detuning from the resonance. This qualitative difference is salient below, as we use the three-level quantum model and compare with the experimental results. Lastly, to convert both experimental measurements and numerical evaluations of the theory model into this form we followed the prescription  in \cite{Bindra1999,Said1992}, inferring from that the associated terms in the expansion of the intensity dependent refractive index in powers of intensity $n_2$, $n_4$ and $n_6$.

%The simplest model of cavity polaritons treats the medium as a Dicke ensemble of two-level excitons coupled corporately to the same cavity field and is sufficient to account for the linear optical properties at all (cavity) couplings. Since the cavity and excitons (here from dye molecules) are resonant at approximately the same complex frequency,  the near-degeneracy leads to the avoided crossing/matter-photon mode mixing generic to polaritons.  In that picture, as the coupling between the Dicke ensemble and the cavity (linear in the dye density) grows,  the splitting between the polaritons does as well, proportional to the square root of the coupling, as expected in an avoided crossing. 

\section{Materials and Methods}
\begin{figure}[htbp]
    \centering
    \includegraphics[width=0.45\textwidth]{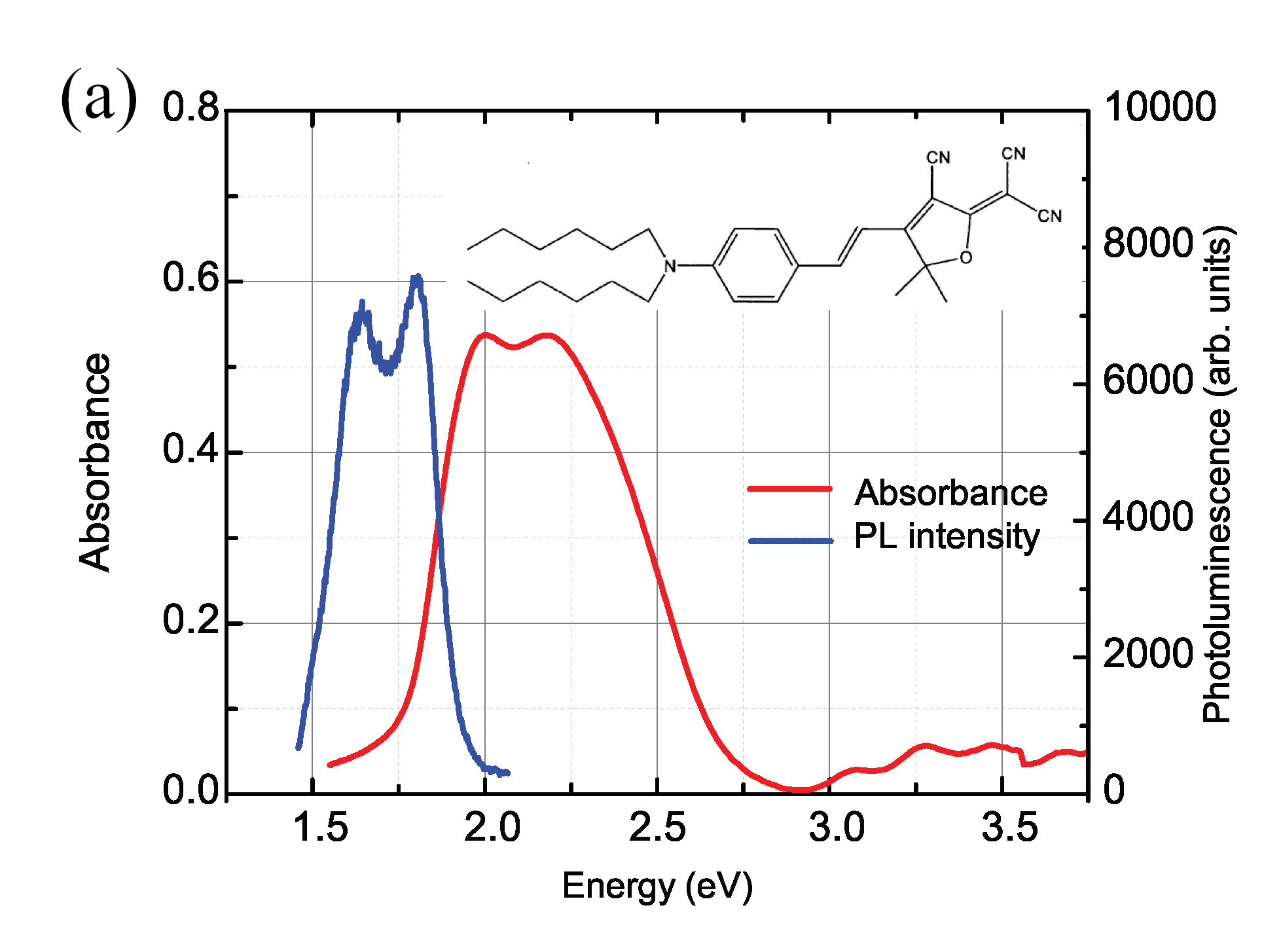}
    \includegraphics[width=0.45\textwidth]{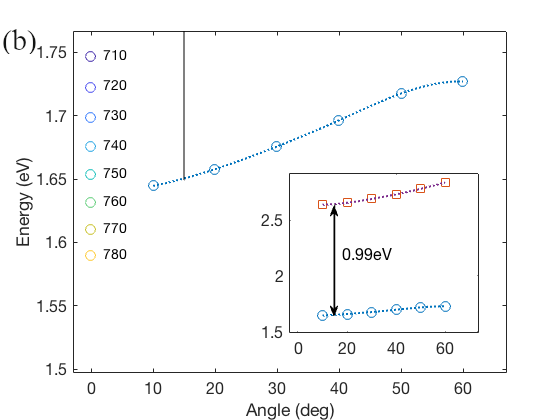}
    \caption{(a) Absorption (red) and photoluminescence (blue) of DCDHF-6V. Chemical structure shown in the inset. (b) Dispersion of polariton sample. Upper (red squares) and LP (blue circles) bands shown as inset, with Rabi splitting energy of about 1eV located at 15 degrees. Inset axes labels are same as main axes. The main figure shows the LP dispersion with the z-scan pump wavelengths from 710nm$\rightarrow$780nm as dark to light circles at normal incidence.}
    \label{fig:DCD}
\end{figure}

The samples were fabricated on glass substrates by first thermally depositing Ag ($\sim$ 20nm) in a vacuum deposition chamber (Angstrom). Then, a solution of the organic dye molecule DCDHF-6V (absorption and photoluminescence shown in Fig. \ref{fig:DCD}a) and PMMA (2:1 mass ratio) was spun coat onto the silver, obtaining a sub-wavelength thickness of $\approx$135nm. The optical cavity was then finished by depositing the final silver layer ($\sim$ 20nm), yielding an overall quality factor for the cavity around 5. The linear optical dispersion derived from reflectivity spectra (SI Fig. \ref{fig:reflectivity}) of the entire structure is shown in Fig. \ref{fig:DCD}b along with the wavelengths associated with the z-scan experiment. Our samples exhibit a Rabi splitting energy of 0.99 eV, placing our samples into the ultra-strong coupling regime.

Experiments were conducted using a parametric amplifier (TOPAS) pumped by a 200 fs laser system (Clark MXR) with repetition rate of 1 kHz. The closed and open aperture z-scan setup is depicted in the supplemental information. The samples were illuminated with average powers of 1, 2, 4, 8, and 12 µW (peak intensity of 0.6, 1.2, 2.4, 4.8, and 7.2 $GW/cm^2$) and their associated open and closed aperture transmissions were recorded using Si large area biased photodetectors (Thorlabs Det100A2). The data were recorded every 50 µm over 20 mm centered around the focus of the Gaussian beam using a translation stage, then centered through calibration with the translation stage in post. The data were averaged over 30 pulses at each spatial point in addition to averaging over four total z-scans.

\section{Experimental Results}

\begin{figure}[htbp]
    \centering
    \includegraphics[width = 0.45\textwidth]{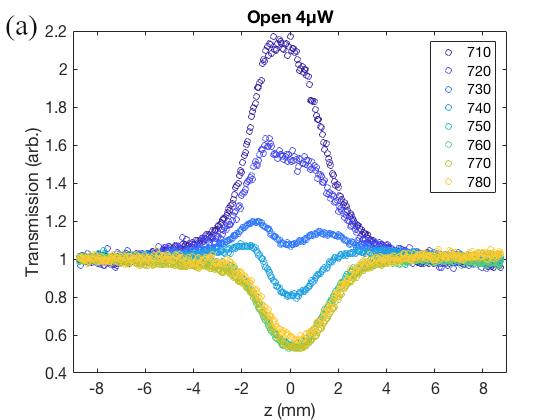}
    \includegraphics[width = 0.45\textwidth]{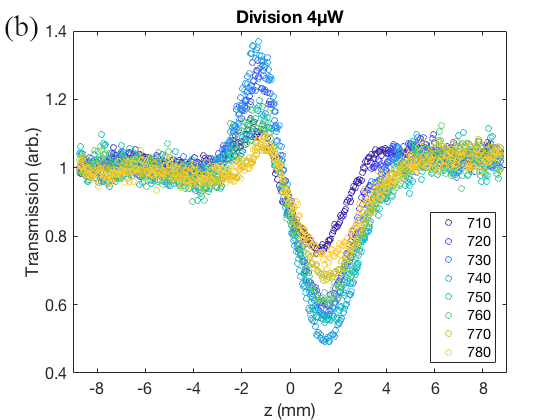}
    \caption{(a) Normal incidence open aperture and (b) closed/open aperture z-scan data scanned at 4 µW average power at each respective wavelength.}
    \label{fig:z-scan}
\end{figure}

Z-scans were taken at wavelengths on both sides of the nominal LP resonance and at various powers. The dominant feature of the open aperture was found to transition from enhanced to diminished transmission from the far-blue to the far-red side of the resonance, but in the intermediate region close to the polariton resonance we consistently measure a re-entrant resonant feature indicating a mixture of the two as shown in Fig. \ref{fig:z-scan}a. This re-entrant resonant feature was present at all power levels, but its visibility for a specific pump wavelength increases with incident power. We  measured a large response in the closed aperture detection arm indicating throughout a negative nonlinear refractive index for our system. To remove the effects of overall intensity change, we divide the closed aperture data and the open aperture data, also shown in Fig. \ref{fig:z-scan}b to obtain the real part of the intensity dependent refractive index. For a baseline  we measured the nonlinear response of a thick film of DCDHF-6V ($\sim$370 nm) (no mirrors and so no cavity polaritons) using the same preparation as the cavity polariton sample. We found the effective nonlinear index, $n_2$, at 680 nm of this cavity-less thick film to be  $1.58\times10^{-16}$ $m^2/W$.  Experimental open and closed/open aperture data for this thick film are included in the supplementary information. 

\begin{figure}[htbp]
    \centering
    \begin{minipage}{0.45\textwidth}
        \includegraphics[width=\linewidth]{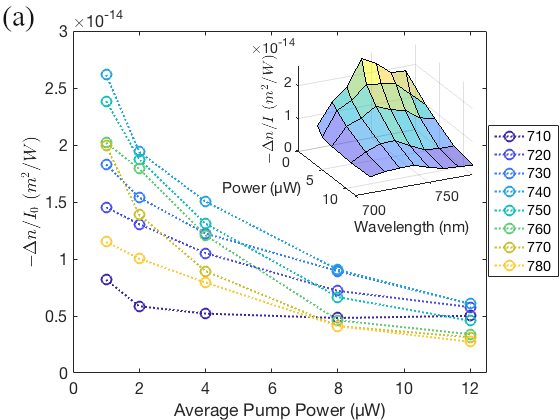}
    \end{minipage}
    \begin{minipage}{0.45\textwidth}
        \includegraphics[width=\linewidth]{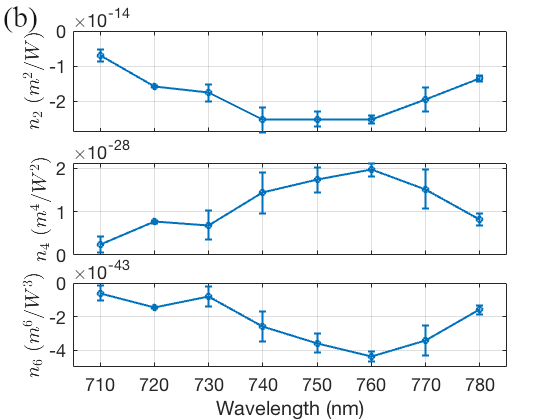}
    \end{minipage}
    
    \caption{(a) Trends with intensity for change in index/intensity, where each line corresponds to a pump wavelength in nm for various input intensities. Inset shows surface plot of same data, but highlights the enhancement around the polariton resonance. (b) Results from fitting process of the data in panel (a) using an order two polynomial fit. These parameters are effective $n_2$, $n_4$, $n_6$ (top to bottom, respectively) of the system.}
    \label{fig:deltan}
\end{figure}

We also investigated the power dependent nature of our responses to determine which high-order processes contribute to our data. Five average power levels were taken for all wavelengths from 710 to 780 nm. The overall transmission difference, from peak to valley, of the z-scan division are shown in the supplemental information, and the extracted $\Delta n/I$ are shown in Fig. \ref{fig:deltan}a. We see a general enhancement of the relative index change around the polariton resonance, but slightly blueshifted by about 10 nm. At higher powers this blueshift becomes larger. The nonlinear enhancement and evidence of higher-order effects is made apparent in the main panel of Fig. \ref{fig:deltan}a. The associated total index change ($\Delta n/I$) becomes diminished at higher power and indicates a high-order effect of opposite sign. It is worth noting that the intensity dependence and curvature in this graphic necessitates the inclusion of both $n_4$ and $n_6$ terms. (see Eq. \ref{dn_overI}) 

The Fig. \ref{fig:deltan}b panel shows the extracted nonlinear indices which result from fitting the $\Delta n/I$ data to an order-two polynomial to match that of Eq. \ref{dn_overI}. We observe up to 150-fold enhancement of $n_2$ relative to the non-polaritonic result noted above. In addition, there is an apparent alternation of overall sign in these extracted indices and a general symmetrical enhancement around the polariton resonance. This also cross-checks with the appearance of the re-entrant feature in the open zscan (Fig. \ref{fig:z-scan}) according to Ref. \cite{Gu2005}, where they assert that it directly implies different signs for the $n_2$ and $n_4$. 

\section{Discussion} 

\noindent \textit{Full Quantum Optics Model}

 Recently, it was shown in \cite{Liu2019} that third harmonic generation into the polariton branches has a dispersive character indicative of the polariton state rather than the exciton. In the sum-over-(intermediate) states picture of third harmonic generation this fact indicates that the natural basis states for perturbation theory is not the exciton but the polaritons themselves. In a similar vein below, we explain the wavelength dependence of our more recent z-scan data also cannot be accommodated by a two level model of the dye exciton. Further we show that at minimum, a three-level model is necessary to connect with the experimental results. 

%\begin{figure}[htbp]
%    \centering
%    \includegraphics[width=0.45\textwidth]{images/3level.png}
%    \caption{Minimal three-level model for dye. 'g' ground state with a large dipole matrix element to the 'e' exciton manifold. 'pl' is the longer-lived stokes shifted image of the exciton manifold. }
%    \label{fig:3level}
%\end{figure}

\begin{figure}[htbp]
    \centering
    \includegraphics[width=0.45\textwidth]{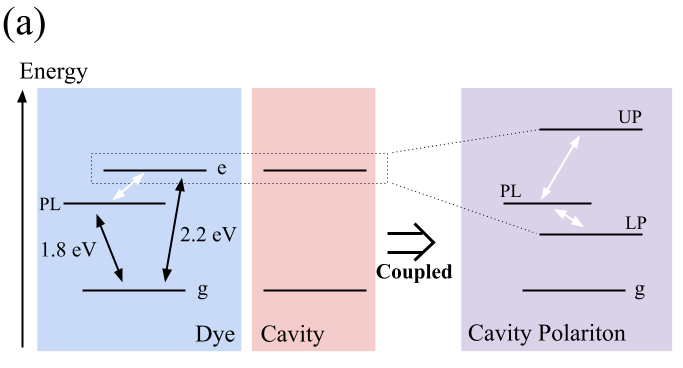}
    \includegraphics[width=0.45\textwidth]{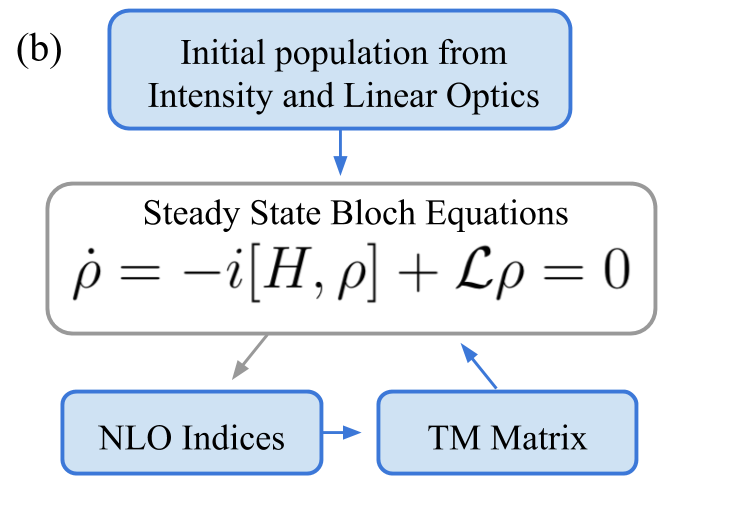}
    \caption{(a) The individual excitons co-operatively couple to the cavity fields. The dispersion in their collective optical response splits the nearby cavity resonance into two states, the upper ("UP") and lower ("LP") polaritons. Black are radiative transitions, white are non-radiative mixing. (b) General flow diagram showing how the numerical evaluation of our polariton system embedded in a z-scan setup takes place. The self consistent steady-state solution emerges from the final three blocks.}
    \label{fig:theoryDescipt}
\end{figure}

We note that, in general, three-level models appear to be minimal for capturing the linear and leading nonlinear phenomenology of most dyes \cite{Dirk92,3levelDye_1, 3levelDye_2}. Taking a cue from the absorption and photoluminescence curves of Fig. \ref{fig:DCD}a, our level scheme for the organic dye, diplayed in Fig. \ref{fig:theoryDescipt}a, consists of a ground state connected by a large matrix element to an excited state centered at 600nm ($\sim$2.1eV, the exciton) and a third state at 660nm (1.8eV, we call "PL", not shown) above the ground state that itself only slowly (radiatively) decays to the ground state. This state is considered since it is accessible during the time characteristic of the Rabi frequency.  Note that the UP state is not included in the model. Further, in addition to the radiative processes between each the exciton or PL and the ground state, we include a fast non-radiative mixing between the PL state and exciton state. This describes significant rapid quenching of the exciton state into the PL state. 

Since the absorption and relaxation processes are fast compared to the excitation pulsewidth (see experimental section), it is sufficient to solve the associated three-level Bloch equations in steady state and use the coherences found in that limit to compute their contribution to the (complex, nonlinear) index of refraction for the dye layer. That index is then used in the transfer matrix modelling of the cavity polariton system. Note that to capture the nonlinear optical response of the system we must include intensity-dependent changes in the index which critically modify (spectral location and depth of) the polariton states  'UP'  and 'LP'.  We do so at each given wavelength and incident intensity by iteratively updating the cavity fields after recalculating the index and transfer matrices until a self-consistent (stationary) cavity intensity is achieved. Diagram  Fig. \ref{fig:theoryDescipt}b is a schematic of this process. 

We numerically evaluate this model to emulate the z-scan experimental protocol. In brief, the z-scan protocol \cite{Sheik-bahae89,VanStryland98} consists of focusing a Gaussian laser field into a waist and translating the sample (the cavity polariton slide) through that waist. Downstream from the waist is a lens for collecting both the total light transmitted through the sample (the "open" z-scan signal) and light that arrives behind the collection lens strictly on axis (the "closed" z-scan signal). The "open" signal indicates nonlinear absorption whereas "closed" channel passing through a nearly closed iris indicates a combination of nonlinear refraction and nonlinear absorption. 

Some details of the numerical modeling are useful for understanding the meaning and limitations of our theory results described below. At each longitudinal sample location, $z$, we break the incident gaussian optical field into hundreds of concentric circular iso-intensity rings and evolve the field in each ring  self-consistently as described earlier. To then form the expected "open" z-scan signal we simply sum the intensities emanating from each ring on the far side of the sample.

To compute an expected closed z-scan signal, at each sample location we use the same self-consistent solution of the transfer matricies to  determine the phase retardation from propagating through the sample at each of the concentric rings individually. We then use the radial map of these phase retardations to determine an effective focal length of the  intensity-induced lensing in the sample. Finally, this is combined with the optical geometry of the downstream part of the experiment (location and focal length of any collection optics) to determine the light intercepted by the iris before the photodetector, yielding the  "closed" (channel) signal.    
\newline

\noindent{\it Comparison to Experiment}

  Much of the observed z-scan data delineating the nonlinear response in ultrastrongly coupled organic cavity polaritons result from a single physical effect: the  optical saturation of the dye. The remaining observations beyond this simple description are experimental evidence for the necessity of including the effects of a third non-radiative level ("PL") in the underlying quantum optics model of the dye. In this section, we first show the result of fixing our one free parameter by matching the experimental blue-shift, then compare the theoretical results of the open and closed aperture response by identifying the key features in both experiment and theory. Lastly, we will show the effective nonlinear response as a function of wavelength and incident intensity, which leads us to a comparison of extracted nonlinear indices as discussed earlier. 

\begin{figure}[htbp]
    \centering
    \includegraphics[width=0.45\textwidth]{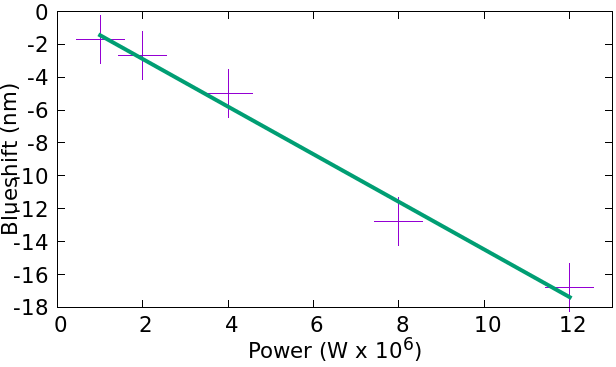}
    \caption{The wavelength at which the open zscan at $z=0$ has a transmission of 1 (matching large $z$)  noticeably blueshifts with the power. Green curve is from evaluation of the theory and the points are experimental data. Note, this is not a fit.}
    \label{fig:centersComparison}
\end{figure}

Our three level quantum optics model has essentially one free parameter: the PL mixing rate. By fixing this we change the rate by which the polaritons may saturate, and thus the blue-shift of the LP (a more detailed description of this is in the following section). In Fig. \ref{fig:centersComparison} we show at various power levels, the  change in wavelength from the low-intensity polariton resonance associated with where the center (z=0) of each normalized z-scan trace transitions through a value of one. This quantifies the blue-shift of our system, and we find strong agreement between the trends in both experiment and theory. More information on this process is discussed in the supplemental information.

\begin{figure}[htbp] 
    \centering
    \includegraphics[width=0.45\textwidth]{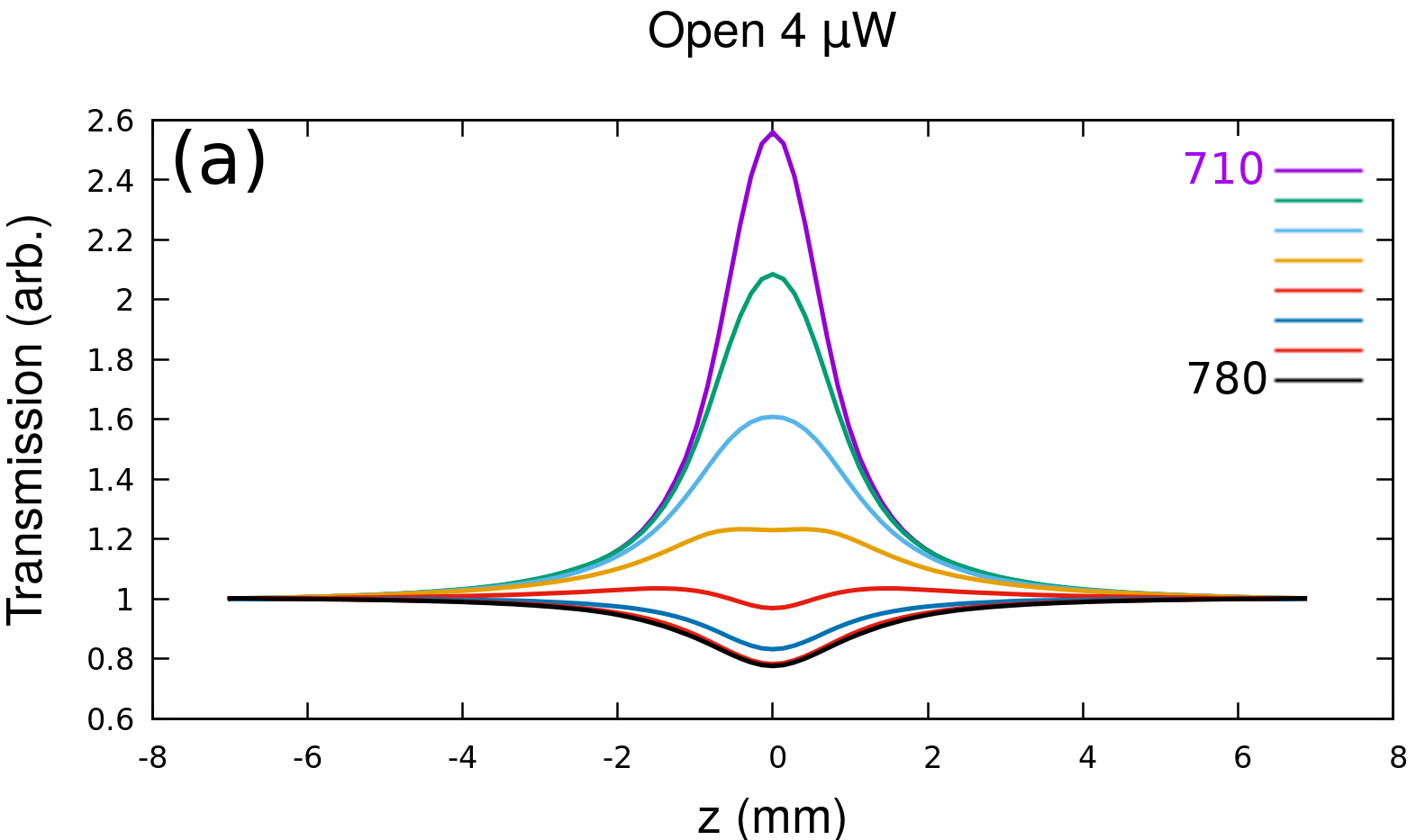}
    \includegraphics[width=0.45\textwidth]{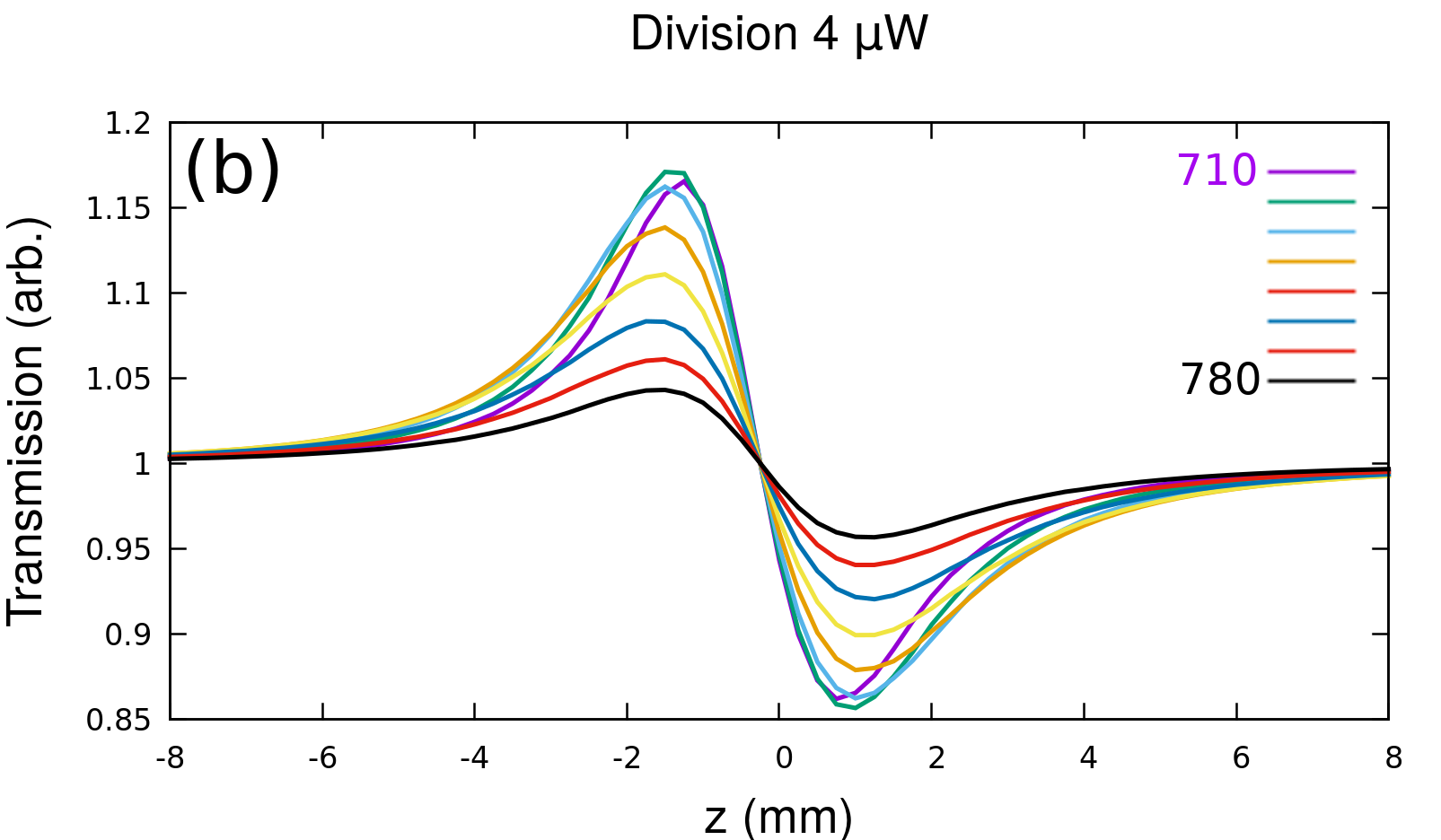}
    \caption{(a) Open zscan theory traces. (b) Closed/Open Zscan theory curves. Legend indicates traces from 710-780nm, each 10nm apart as in Fig. \ref{fig:z-scan}. Mixing rate set to 4 x 10$^5$ GHz between the exciton and the "PL" state. Compare with experimental Fig. \ref{fig:z-scan}, though these theory simulations used 4 $\mu$W of optical power.}
    
    \label{fig:theoryOpenClosed}
\end{figure}

By fixing this mixing rate, we can evaluate the full model and compare the open and closed/open channels in both theory and experiment. Shown in Fig. \ref{fig:theoryOpenClosed} is the result of our three level model. We take note here to draw the qualitative comparison to the experiment in Fig. \ref{fig:z-scan} in both the open and division data. There are 3 main qualitative features in the open aperture data and 2 main features in the division data we will compare. We will also discuss how they are the result of optical saturation for the open z-scan data for these ultrastrongly coupled organic cavity polaritons.  

1) SA- to RSA-like transition: As seen in Fig. \ref{fig:z-scan} and \ref{fig:theoryOpenClosed}, scanning in wavelength across the polariton resonance(s), the open z-scan data change by processes similar to saturable absorption (SA-like) (reduced nonlinear absorption) to reverse saturable absorption (RSA-like) (increased nonlinear absorption). The wavelength at which this transition occurs shifts blue with increasing intensity, which follows the same trend as depicted in Fig. \ref{fig:centersComparison}. 

Saturation due to the brightening of the internal cavity optical fields reduces the cavity coupling. This reduces the vacuum Rabi frequency, reducing the gap between the polariton resonances.\cite{gap_closing} This causes the UP to move to longer wavelengths and the LP to shorter ('blueshift'). Theory model evaluations of this are included in the supplemental information Fig. \ref{fig:mind_the_gap}. At low intensities (linear optical regime), the polariton resonances increase the intracavity intensity. Thus, if blue detuned from the LP, then during the intensity increase in a z-scan the LP moves towards the drive wavelength, resulting in SA-type behavior in the open z-scan channel. The same blueshift causes an intensity decrease when the LP is scanned at a red detuned wavelength, thus appearing RSA-like. If saturation dominates we would expect exactly the opposite behavior crossing the UP, which indeed is indicated by our experimental findings (not shown here). Note these findings cannot be reproduced by a simple wavelength-independent nonlinear index for the dye, because in that case both UP and LP would shift the same sense with intensity, as indicated in Eq. \ref{Deltan}

2) Re-entrant ("M"-shaped) feature in middle of SA-RSA transition: Also apparent from Fig. \ref{fig:z-scan} is that the open aperture signal has a re-entrant behavior at wavelengths in the transitions from the SA- RSA-like ranges. With increasing optical power this re-entrant transition feature for the LP blueshifts in wavelength, broadens about $z$, and increases in contrast. 

As a result of the blueshift of the LP with intracavity intensity, the LP resonance center shifts so far that it passes to the blue of the excitation wavelength itself. In that case the absorption first decreases at $z\neq0$ as the blueshift pulls the LP resonance towards it and then appears to increase as $z\rightarrow 0$ since there it continues to blueshift the LP so far that at that wavelength the scan goes off resonance again {\it on the long wavelength side}. We note that there is a rich literature of such transition features \cite{m1, m2,m3,m4,m5,m6,m7} in open z-scan data, and in all cases the feature is a consequence of intensity dependent frequency pulling of an optical resonance of some sort. This picture also explains the increased broadening in z-scan co-ordinate, increased contrast with power, and implies that the same phenomenology the re-entrant transition feature should appear red of the UP, which we have also verified experimentally (not included here). 

\begin{figure}[htbp]
    \centering
    \includegraphics[width=0.45\textwidth]{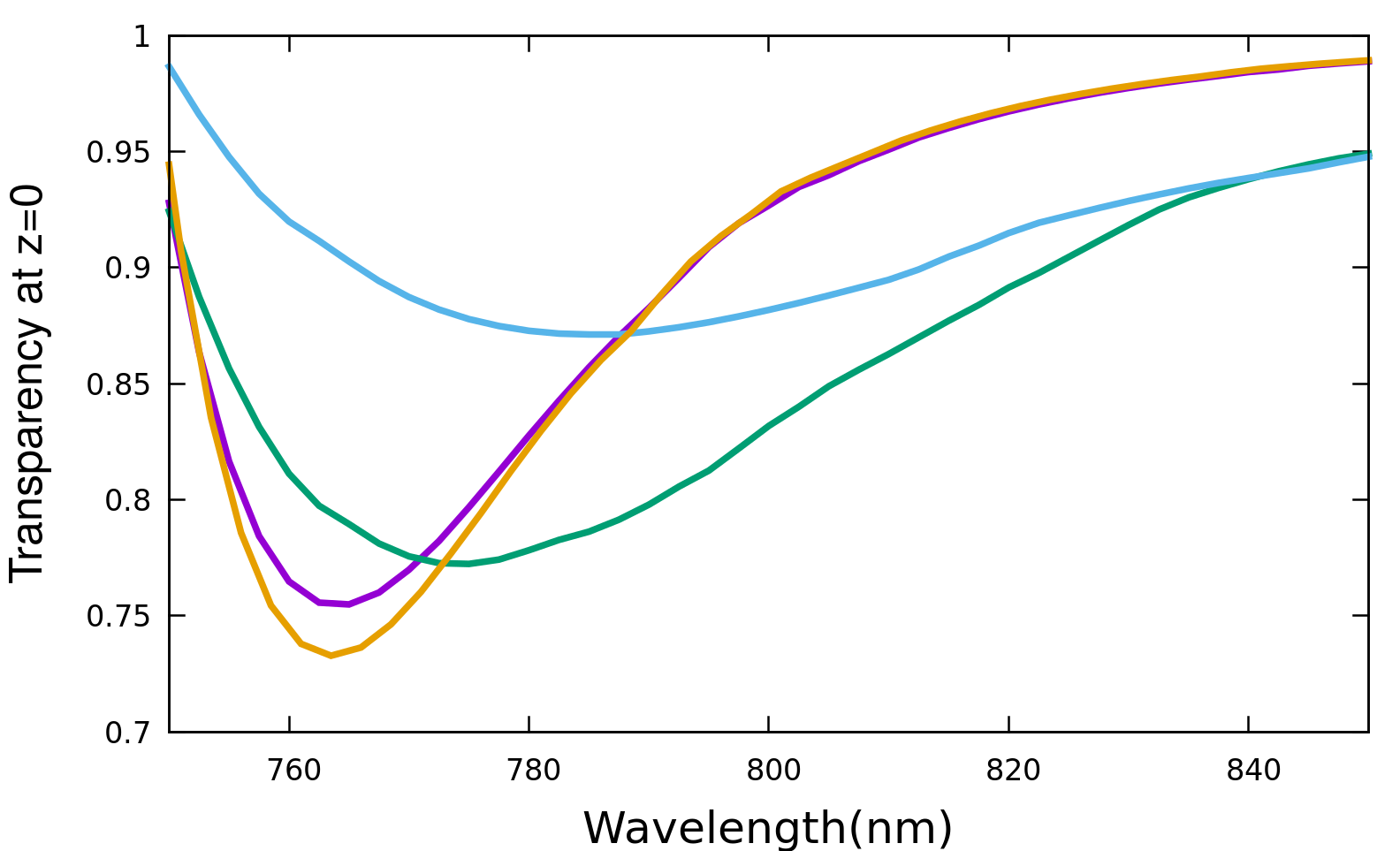}
    \caption{Theory open zscan minima as a function of wavelength, all parameters the same except the mixing rate between the exciton and the "PL" state as described in the text. Power is 4 $\mu$W for all curves. The purple curve is the 2 state model (no mixing) the green is for a mixing rate of 4 x 10$^5$ GHz and the light blue curve is with a mixing rate of 8 x 10$^5$ GHz. The gold curve is the square of the two state (no mixing) model at half the power, showing that changing the power cannot explain the observed persistence of the additional nonlinear absorption red of the LP.}  %Points are from our experiment.}
    \label{fig:openTheoryBottoms}
\end{figure}

3) The RSA-like behavior in the open aperture data of the LP persists far into red detuning, to nearly 100 nm beyond the LP. 
This appears to not be a simple consequence of saturation; instead, we understand this behavior as consequent to the active participation of a third level in the quantum optics model of the dye. The participation of this third level through its rapid mixing with the exciton furnishes a longer-lived set of metastable states that do not themselves directly lead to cavity pulling, but through excited state mixing, broaden the frequency response of the dye at high intensity. We directly, quantitatively compare a 2-level (no mixing) and 3-level model (with mixing) for the dye cavity polaritons in Fig. \ref{fig:openTheoryBottoms} by comparing open z-scan minima. 

As described in the supplmentary information, the laser power, beam profile, z-scan optics chromatism and detection chain were well characterized and those measured parameters were used in connecting theory outputs (self-consistent nonlinear optical transfer matrices) to the associated open- and closed-signal channels. Also described there are how all but one of the relevant microphysical parameters of the quantum optics model of the dye are fixed by the linear optical behavior and dye density. Thus the non-Hermitian mixing rate between the PL and the exciton state is the only adjustable parameter in the model.  

Plotting the minimum of the theory open z-scan (i.e. at $z=0$) as a function of detuning from the LP for different choices of that mixing rate (see Fig. \ref{fig:openTheoryBottoms}) allows a quantitative comparison with experiment. 
We therefore adopt a mixing rate of 4.0x10$^5$ GHz in all the numerical evaluations of the theory model here except where noted otherwise. This provides evidence for the necessity of a three-level quantum optics model as minimal for understanding the relevant contributions from the dye nonlinear optical properties, seen here in its NLO effect of the associated ultrastrongly coupled cavity polaritons.  

Having related each observed open z-scan feature qualitatively to the dye quantum optics model, we now use it in the qualitative explanation of the closed z-scan signal, by focusing on two defining features.
%
% Hi Sam!! taking a read of it from front to back now...
%
1) The sign of $\Delta n$, the overall intensity dependent refractive index: The observed closed z-scan signal from the cavity LP correspond to $\Delta n<0$ for all wavelengths and powers. The magnitude of the effect increases as one approaches the blueshifted LP center wavelength, and broadens (in z-scan coordinate $z$) with power.

This is a consequence of the fact that the LP is below the exciton. Even in a two level system, again as a consequence of power broadening/saturation, the first contribution to $\Delta n$ is expected to follow that of normal dispersion, being negative at detunings below the exciton and positive above, as indicated by our experimental findings at the UP (not included here). The broadening seen is consistent with the expected power dependence of the nonlinear response, and we have already shown that its observed blueshift with power is quantitatively consistent with that due to the saturation depolarization of the dye in the cavity optical field. 

2) The z-scan shape changes and dynamic range of the closed/open z-scan with power: As noted, $\Delta n$ from the closed z-scan stays negative across the LP, but its magnitude changes with power differently at various detunings. 

The sign of $\Delta n$ does not tell the whole story. The magnitude of the $\Delta n$ depends on detuning and fluence in such a way as to indicate the vital contribution of higher order nonlinear susceptibilities. As described in the experimental section (and as rendered from theory evaluation in Fig. \ref{fig:theoryDT} below), these changes can be recorded as nonlinear contributions to the index, $n_i$, $i = 2,4,6$.  Qualitatively all three of these show dependence on the detuning (from the polariton) that is a consequence of a blueshift, as indicated qualitatively by expanding out the expression for $n_i$ from Eq.\ref{twoLevel_n}. When we evaluate our model using the same methods as described in the experimental section for extracting $\Delta n/I$, we find significant similarity. Namely, an enhanced response around the polariton resonance, and a change of sign for each higher-order effective nonlinear index, as seen in Fig. \ref{fig:theoryDT} and in the data of Fig. \ref{fig:deltan}b. The general broadening of the response movement to the blue at higher intensities is consistent with what was observed in the experiment, as is the significant agreement in the overall magnitude of the nonlinear response.

\begin{figure}[htbp]
    \centering
    \includegraphics[width=0.45\textwidth]{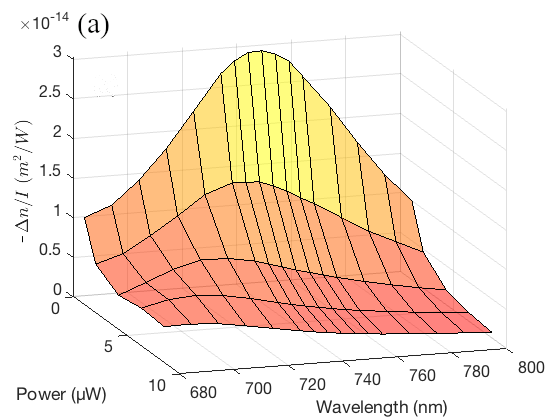}
    \includegraphics[width = 0.45\textwidth]{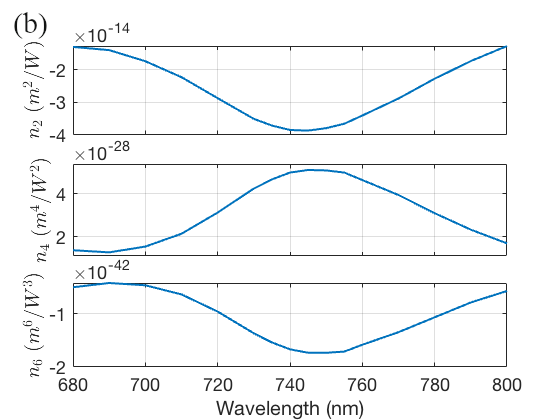}
    \caption{(a) The theory-derived $\Delta n/I$  traces from theory closed/open theory.  Parameters same as in Fig. \ref{fig:theoryOpenClosed}. For (b) we converted the data in (a) into the intensity dependent contributions to the overall index of refraction of the sample. Compare with experimental Fig. \ref{fig:deltan}. }
    \label{fig:theoryDT}
\end{figure}

%\section{Discussion}
\section{Summary and Outlook}
We have carried out experimental and theoretical studies of the nonlinear optical response spectrum of ultrastrongly coupled organic cavity polaritons.  We find up to 150-fold enhancement of the response compared to cavity-less films.   Our experimental findings and their accompanied theoretical elucidation using a straightforward three-level quantum optics model, with essentially only a single adjustable parameter, indicates that for ultrastrongly coupled organic polaritons, the nonlinear refractive index is dominated by one main effect: the reduction of the vacuum Rabi frequency due to the saturation depolarization of the medium in the cavity's intense optical field. This reduction of Rabi frequency produces a blueshift of the LP yielding a complex nonlinear refractive index exhibiting contributions from higher order contributions to $\Delta n$ dependent on the detuning and power broadening as we show with a simple two-level model.  Although the reduction of the Rabi frequency due to saturation with intensity would also readily occur were there only two contributing levels, we find that the experimentally measured behavior of the open aperture behavior at long wavelengths further from resonance (longer than the LP) requires the inclusion of a third level in our quantum description of the dye.

Although that later point is not necessarily surprising, here we have done more by actually qualitatively and quantitatively connecting the underlying microphysical sources of the nonlinearities to those of the more complicated optical geometry. Further work is underway to delineate how dye-instrinsic higher-order nonlinear optical processes contribute to the observed z-scan signals. 

Beyond being simply explanatory, we can use this model and understanding as a tool to predict and manipulate the nonlinear properties of multi-polariton systems, which may be of practical utility for optical switching and quantum information processing using polaritonic matter.   

\section*{Funding}
This work was supported by in part by the U.S. National Science Foundation (Grant DMR-1609077). 

\section*{Acknowledgments}
The  authors acknowledge the use of the Materials for Opto/Electronics Research and Education Center (MORE) for sample preparation  and characterization  at Case Western Reserve University.

\section*{Disclosures}
None of the authors of this paper has a financial or personal relationship with other people or organizations that could inappropriately influence or bias the content of the paper.

%%%%%%%%%% If using BibTeX:
\bibliography{MAIN}

\newpage
\section*{Supplementary Material}

Contained here are the additional graphics and information associated with the main paper. 
\newline

\noindent {\it Reflectivity spectra} 

\begin{figure}[h]
    \centering
    \includegraphics[width=0.45\textwidth]{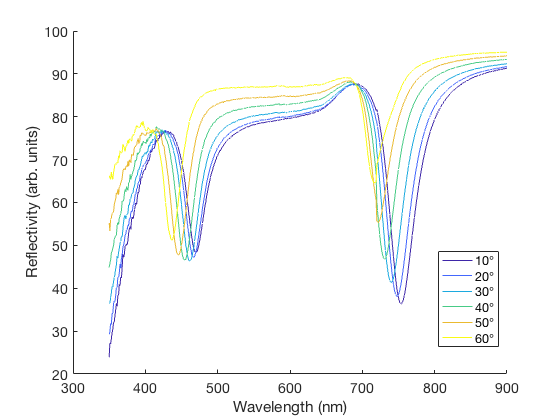}
    \caption{Reflectivity data of the polaritonic sample at various input wavelengths and angles.}
    \label{fig:reflectivity}
\end{figure}

Reflectivity spectra underlying the linear dispersion of Fig. \ref{fig:DCD}b taken with angle resolved Varian Cary 6000i UV-VIS spectrophotometer at 10 degree increments and displayed in Fig. \ref{fig:reflectivity}.\newline

\begin{figure}[h]
    \centering
    \includegraphics[width=0.5\textwidth]{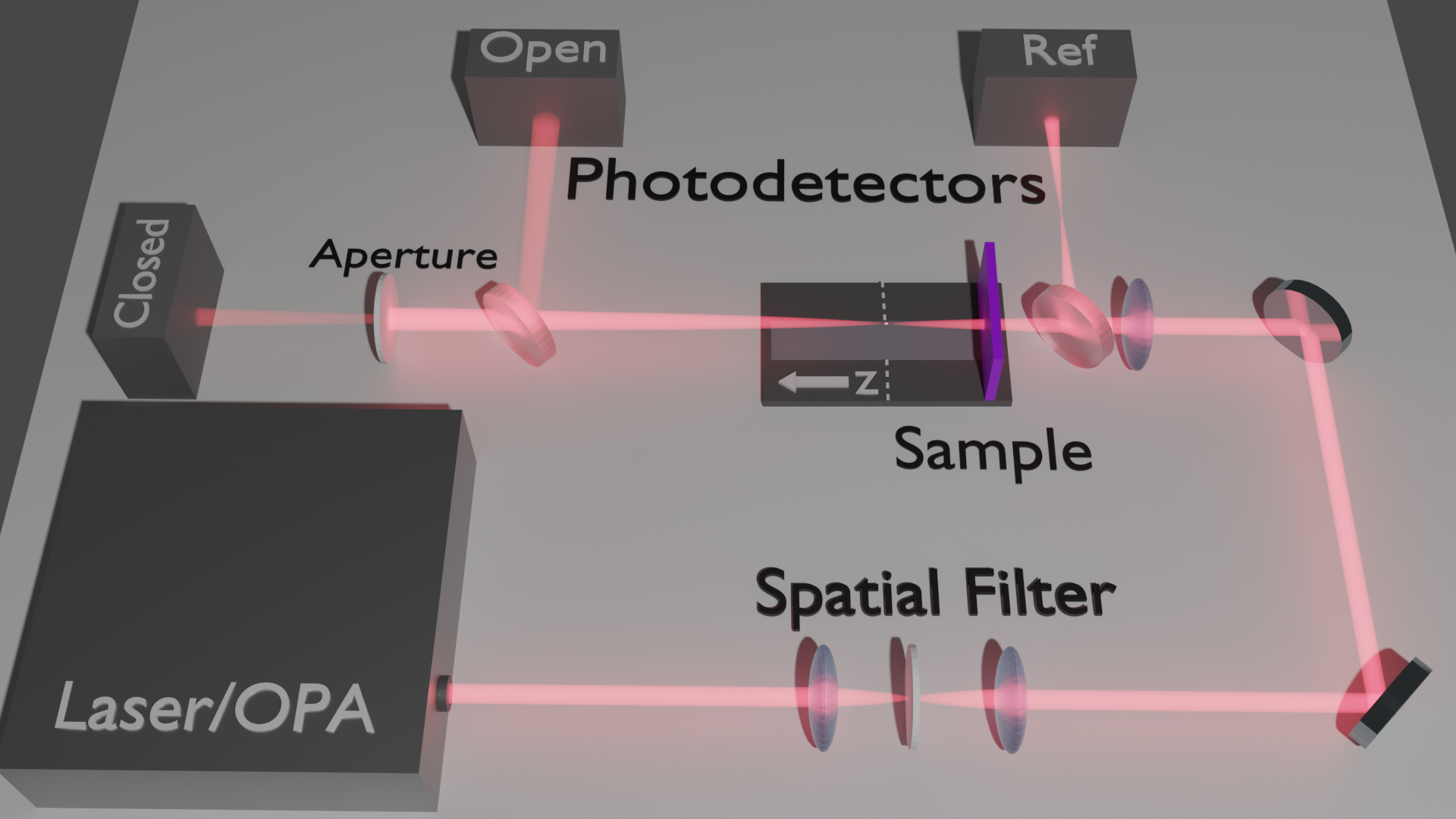}
    \caption{Experimental z-scan setup for open and closed aperture studies. Both arms of detection are recorded simultaneously by using a beam-splitter at the far-field output of the sample. }
    \label{fig:zscan_setup}
\end{figure}

\noindent {\it Experimental setup}\newline
Fig. \ref{fig:zscan_setup} is the diagrammatic representation of our experimental setup as described in the main text. It is a standard z-scan setup with two arms to simultaneously record both the open and closed aperture data from the sample. \newline

\noindent \textit{Two level model}\newline
As discussed in the main text, here we adopt an illustrative 2-level model expansion to display the effects of an intensity dependent blueshift ($\tau$) embedded in the detuning ($\delta$) of this toy system. The two-level model expansion for the nonlinear parameters in terms of our generalized loss, $\Gamma$, and power broadening , $P$, is given below.

\begin{align*}
    &n_2=U(G,\omega)\frac{(\Gamma^2 -\delta^2 )\tau-\delta P}{(\delta^2 + \Gamma^2)^2}\\
    &n_4=U(G,\omega)\frac{\delta  \tau ^2 \left(\delta ^2-3 \Gamma ^2\right)+\delta  P^2-P \tau  \left(\Gamma ^2-3 \delta ^2\right)}{\left(\Gamma ^2+\delta ^2\right)^3}\\
    &n_6=U(G,\omega)\frac{-\tau ^3 \left(\Gamma ^4-6 \Gamma ^2 \delta ^2+\delta ^4\right)-\delta  P^3+P^2 \tau  \left(\Gamma ^2-5 \delta ^2\right)+6 \delta  P \tau ^2 \left(\Gamma ^2-\delta ^2\right)}{\left(\Gamma ^2+\delta ^2\right)^4}
\end{align*}

\noindent which for the limit there is no blueshift  ($\tau = 0$) we find the index is comprised of  co-resonant nonlinear terms of alternating sign, all vanishing on resonance; 

\begin{equation*}
    n-1=\frac{\delta }{\Gamma ^2+\delta ^2} -\frac{\delta  P}{\left(\Gamma ^2+\delta ^2\right)^2}I  +\frac{\delta  P^2}{\left(\Gamma ^2+\delta ^2\right)^3}I^2-\frac{\delta  P^3}{\left(\Gamma ^2+\delta ^2\right)^4}I^3 
\end{equation*}

\begin{figure}[htbp]
    \centering
    \includegraphics[width=0.55\textwidth]{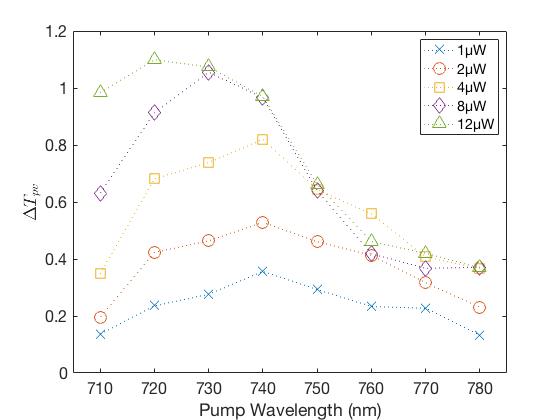}
    \caption{Experimental change in transmission in normalized closed z-scan of cavity polariton sample, peak to valley, for various average power levels.}
    \label{fig:deltaT}
\end{figure}

 This simple, illustrative model indicates why at relatively high incident intensities the largest change in transmission (and thus the largest change in index) can shift in comparison with the low intensity regime. When the blueshift parameter $\tau$ is not zero the different non-linear orders no longer need be alternating in sign or vanish on resonance. 
In our system at the LP we measured data displayed in Fig. \ref{fig:deltaT}. The linewidth and location of the response changes with incident intensity, as one would expect if there were an intensity dependent blueshift, in our case resulting from optical saturation effectively reducing the vacuum Rabi splitting. \newline

\begin{figure}[h]
    \centering
    \includegraphics[width=0.45\textwidth]{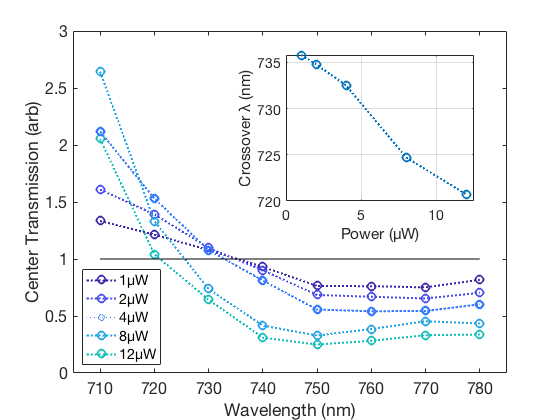}
    \includegraphics[width=0.45\textwidth]{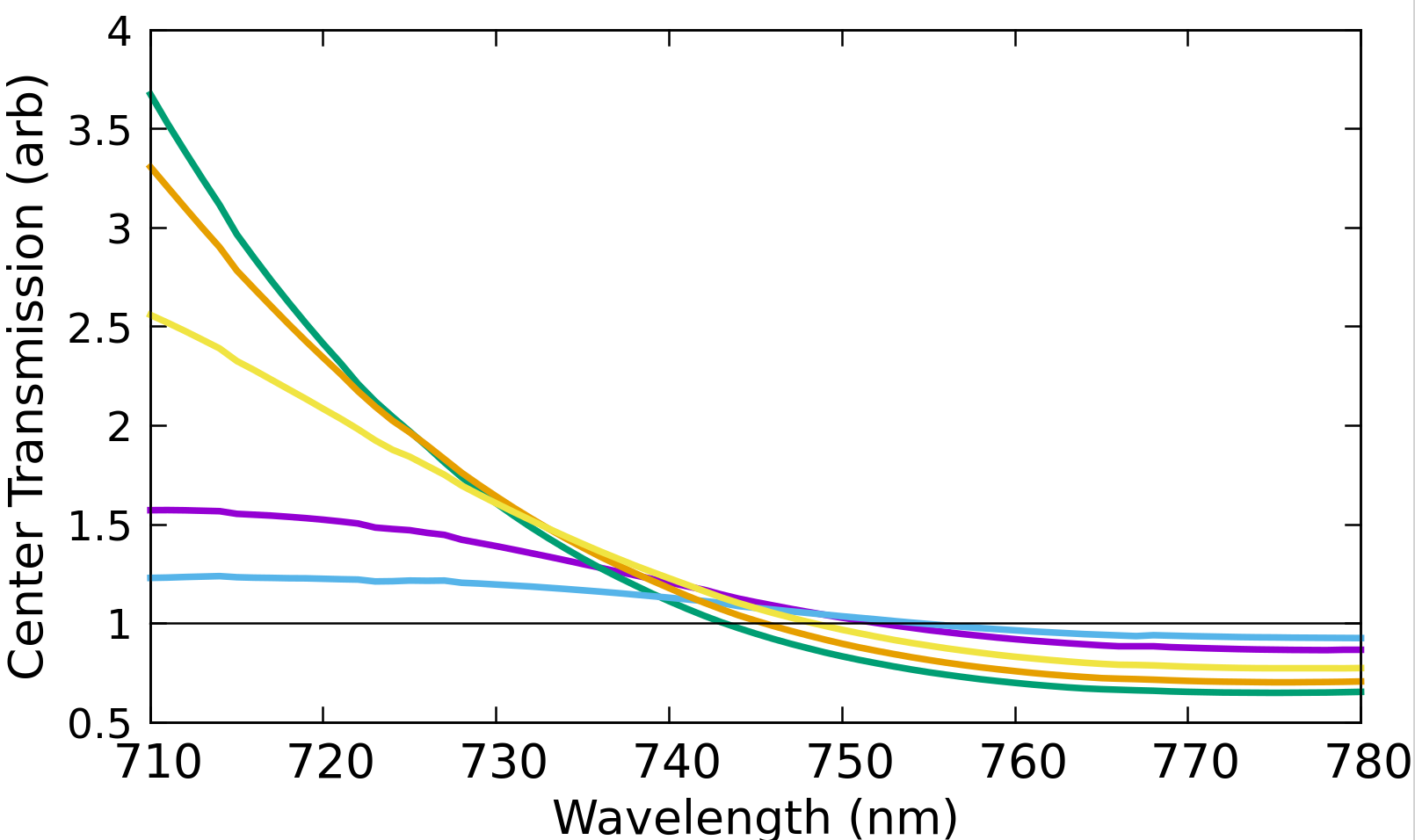}
    \caption{(a) Center of experimental z-scan for each power level investigated as a function of pump wavelength. (b) Theory for powers 1, 2.1, 4.3, 6.4, and 8.6 microwatts (blue to green, resp.) power}
    \label{fig:centersPlot}
\end{figure}

\noindent \textit{Three level model}\newline
We now discuss the origin for the data populating Fig. \ref{fig:centersComparison}, which displays the blue shift power dependent trends of both the theory and experiment. For these data we study the center (z=0) of each z-scan for each power level. This transmission value is recorded for every wavelength that was used in the experiment and theoretical modeling, summarized in Fig. \ref{fig:centersPlot}. Each trace shows the evolution of the SA- to RSA-like transition from the blue$\rightarrow$red side of the polariton resonance, as discussed in the main text. The point where these traces cross 1 is the corresponding data found in the main text Fig. \ref{fig:centersComparison}. The inset of the left panel of Fig. \ref{fig:centersPlot} shows the experimental crossing of the threshold, reproduced also as the points in Fig. \ref{fig:centersComparison}.

\begin{figure}[h]
    \centering
    \includegraphics[width=0.45\textwidth]{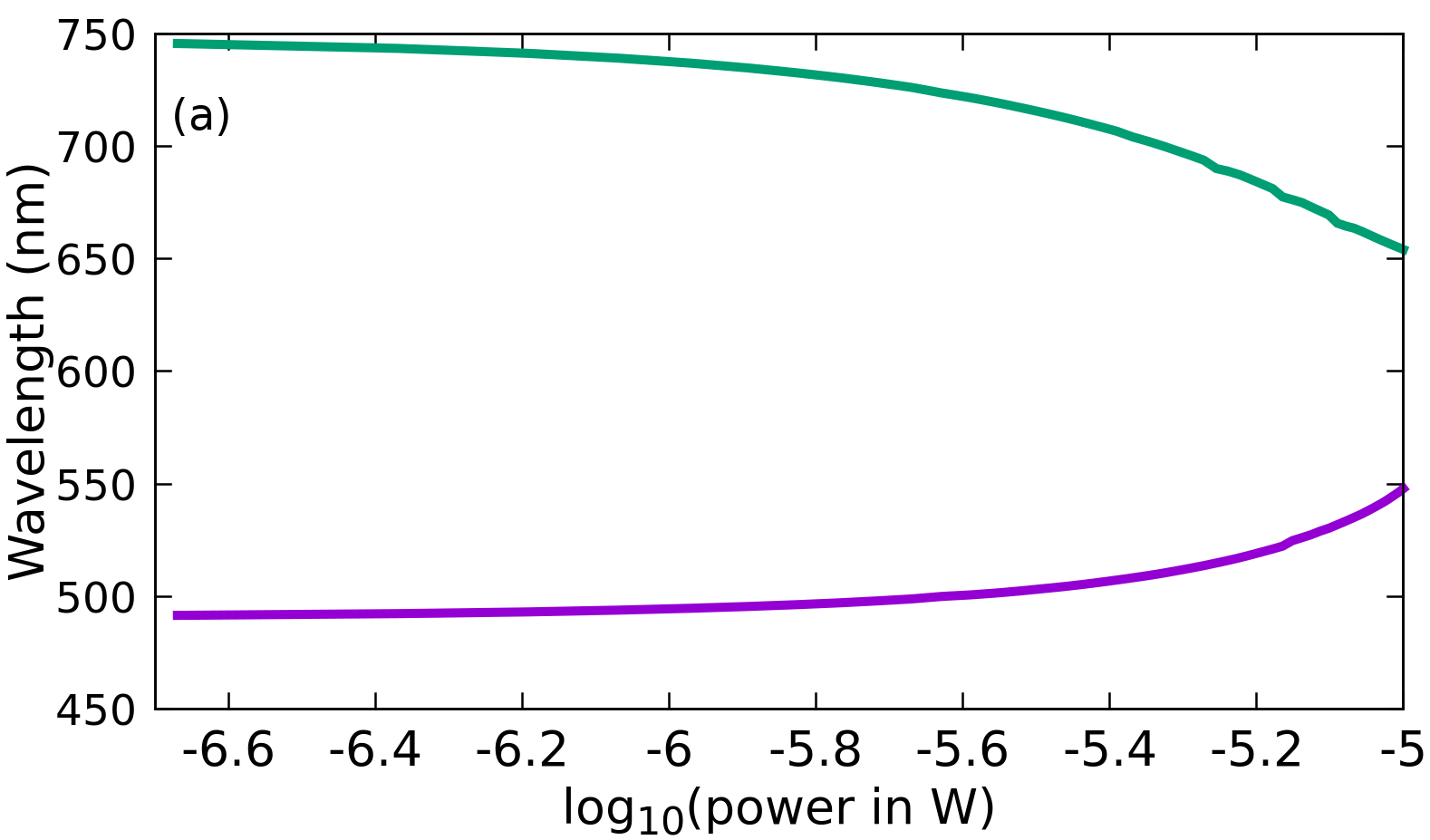}
    \includegraphics[width = 0.45\textwidth]{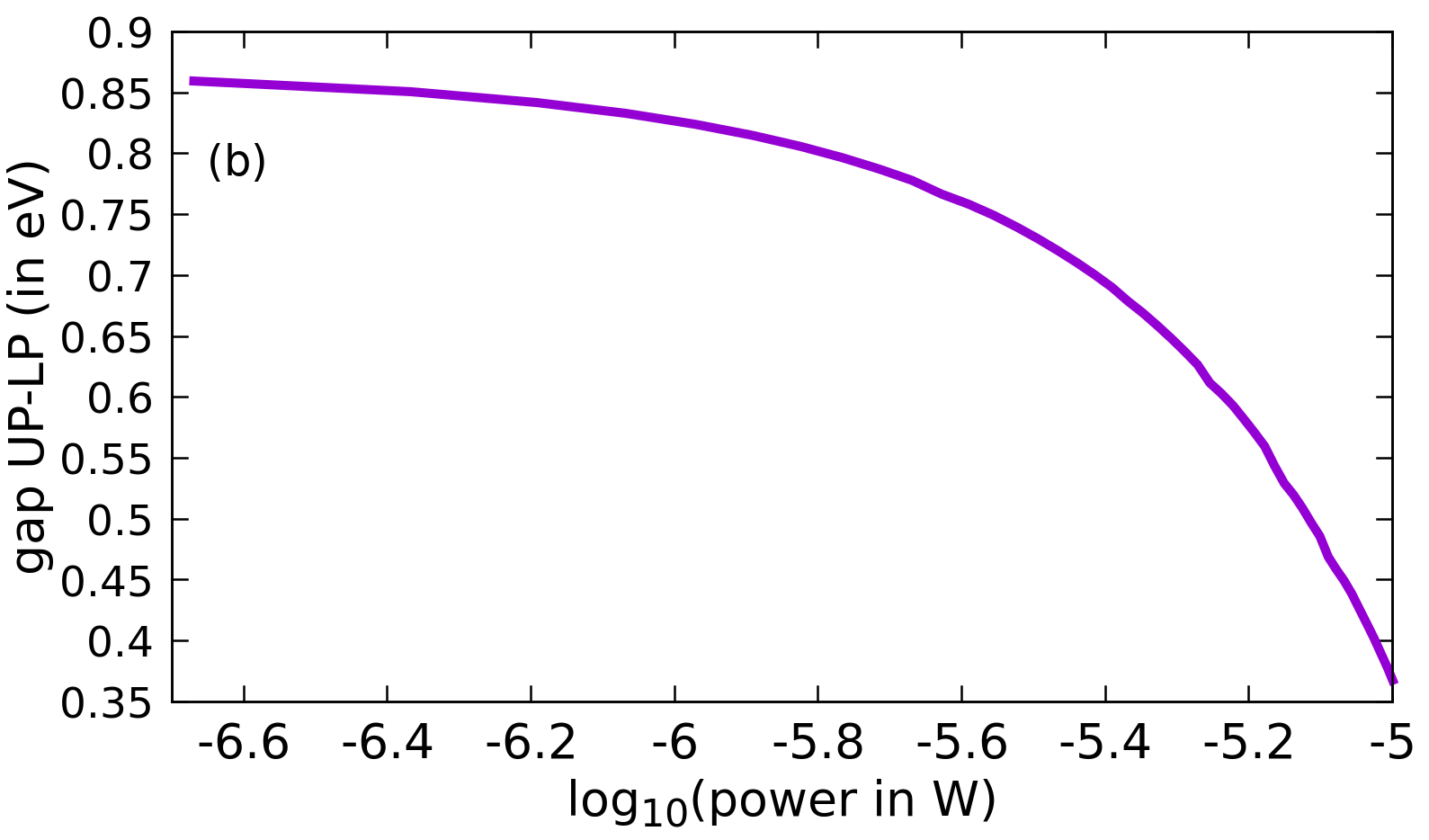}
    \caption{The (a) change of the individual polariton branches for a sample at $z=0$ and (b) rendered in terms of the polariton gap in eV as a function of the pump power, from evaluation of the quantum optics model and transfer matrix transport.}
    \label{fig:mind_the_gap}
\end{figure}

As further illustration of the mechanism associated with the observe blue shift of the lower polariton, we display theory results of increasing the optical intensity incident on the polariton sample. In Fig. \ref{fig:mind_the_gap}
we see a clear Rabi bleaching as the intensity increases.\cite{gap_closing} Thus, if on the blue side of the LP during a (open) zscan one experiences an increase in the overall transmission from a shift in the linear optical response, with exactly the opposite behavior expected at the UP. This qualitative difference was observed experimentally using our cavity polariton sample. 

\begin{figure}[h]
    \centering
    \includegraphics[width=0.45\textwidth]{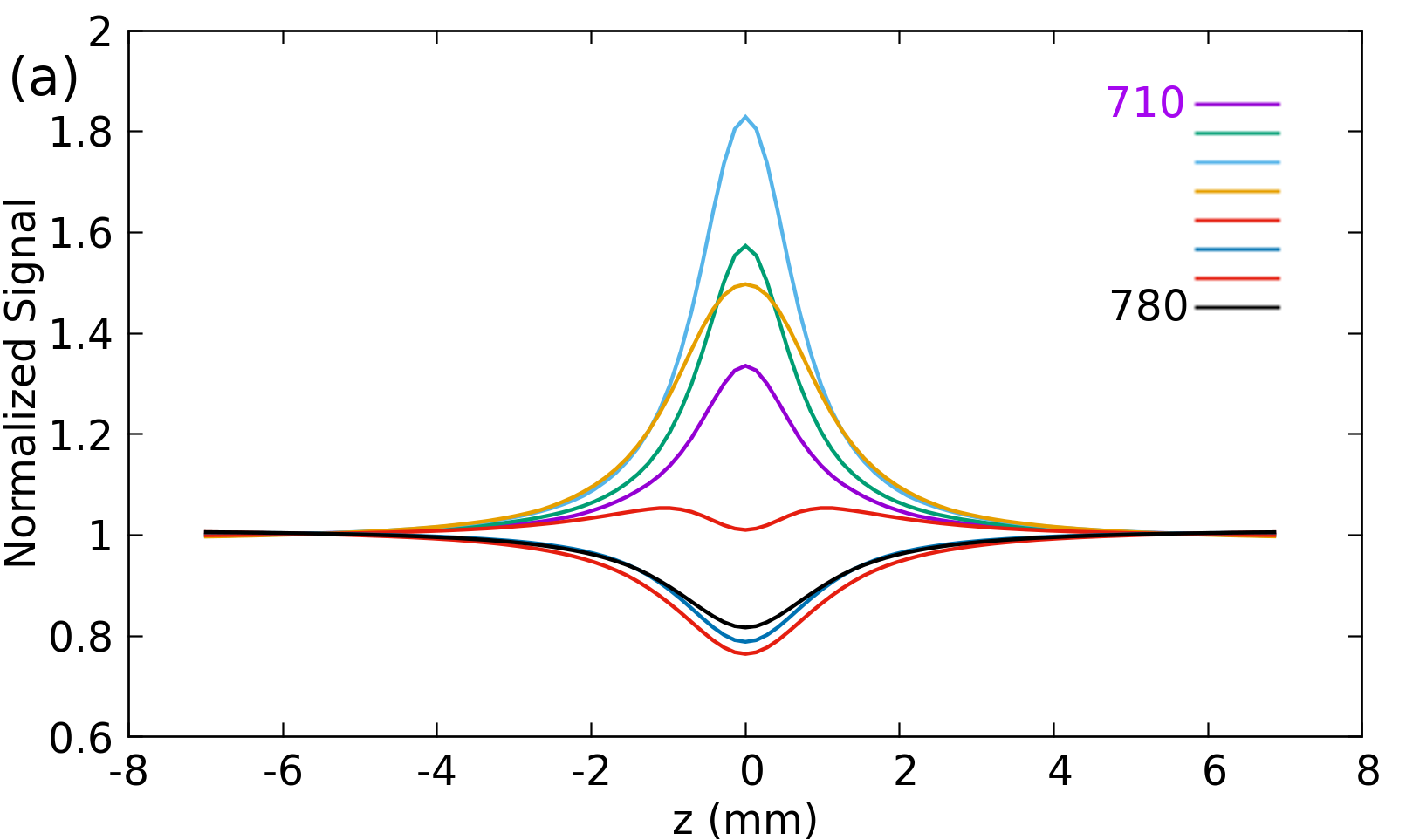}
    \includegraphics[width=0.45\textwidth]{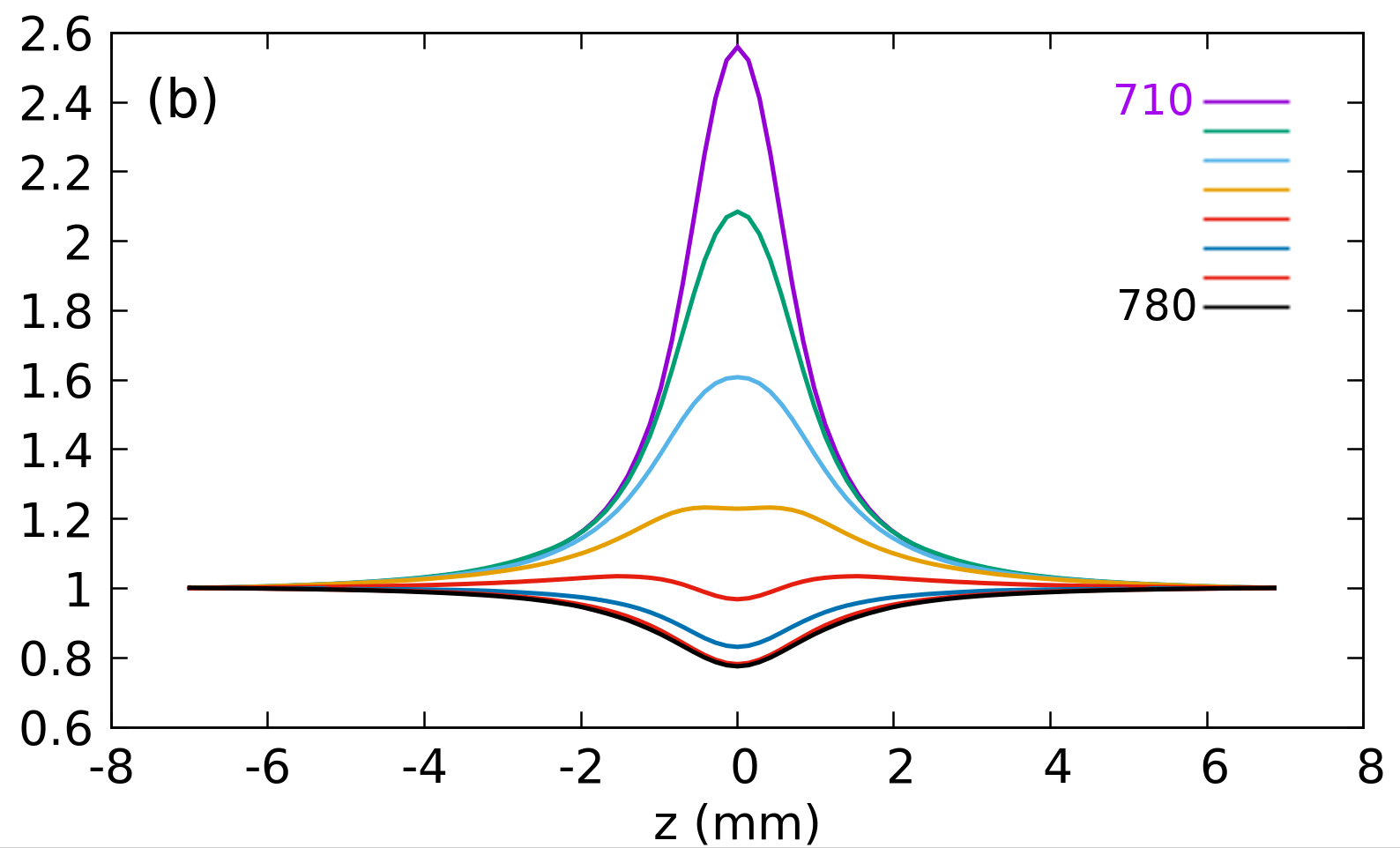}
    \caption{Open zscan theory traces (710-780nm , each 10nm apart as in Fig. \ref{fig:z-scan}) for (a) small mixing (essentially a two state model for the dye) and (b) large mixing rate (=4 x 10$^5$ GHz between the exciton and the "PL" state), all other parameters the same. 
    Compare with experimental Fig. \ref{fig:z-scan}(a), though these theory simulations used 4 $\mu$W of optical power.}
    \label{fig:openTheory}
\end{figure}

We also briefly discuss the implications of the 3 level model and the importance of our final free parameter, the mixing rate. As shown in open zscans of Fig. \ref{fig:openTheory} there is a significant difference between a simple two level model (small mixing rate) and the three level model. For large detunings the nonlinear overall response persists when a third level is included. Without that third level the response more quickly returns back to normalized transmission = 1 for large detunings from the polariton resonance. \newline

\noindent {\it Model Inputs:} 

We now briefly review the parameter inputs we use to numerically evaluate the model. Every parameter but one (the mixing rate between the exciton and the PL states) is essentially fixed by the known geometry of the setup, measured powers, wavelengths and beam sizes and the dye's linear optical response. 

It is convenient to separate the parameters into microscopic and macroscopic. The microscopic parameters refer to the inputs to the quantum optics model. As described earlier, they consist of exciton line center and effective width (measured for this dye in neat films by ellipsometry; see Ref. \cite{Liu2015} about 600 nm line center and 1.5e4 GHz width), the gap (Stokes shift) between that state and the PL state, here fixed at 0.3eV below the exciton) and finally the excitonic dipole matrix element (also from Ref. \cite{Liu2015}, measured to be about 7.3 D). The single important unknown microscopic parameter we can vary theoretically is the mixing rate between the exciton and the PL state. While there are potentially other microphysical parameters in the model (the decay rate of the PL state, the branching ratio of decays, the coherence decay rate between the excitonic state and the PL state, etc.) we find in fact that varying these other parameters (over reasonable ranges) does not materially effect the quantitative results. 

All macroscopic parameters are also  essentially fixed by linear optical measurements and well-characterized sample fabrication protocols. These include the layer thicknesses (deposition process control leaves those with variances below the 10\% level) of the silver mirrors and the dye layer, the dye concentration with PMMA carrier, etc. The simple quantum optics model as described combined with the known dye density reproduces the (complex) linear optical index measured via ellipsometry, again, at the 10\% level across the visible. 

To model the optical response of the deposited silver mirrors we used the published bulk values of $n+iK$ of Ref. \cite{McPeak}. 
We take throughout this manuscript our front and back silver layers to be 18 and 20 nm thick (resp.) and the dye layer to be 140 nm thick and the dye molecular number density of $\sim$ 1.6 x 10$^{21}$/cc, which we note is consistent with the chemical preparation, but also proscribed by the measured vacuum Rabi splitting of the polaritons. 
\newline

\begin{figure}[htbp]
    \centering
    \includegraphics[width=0.45\textwidth]{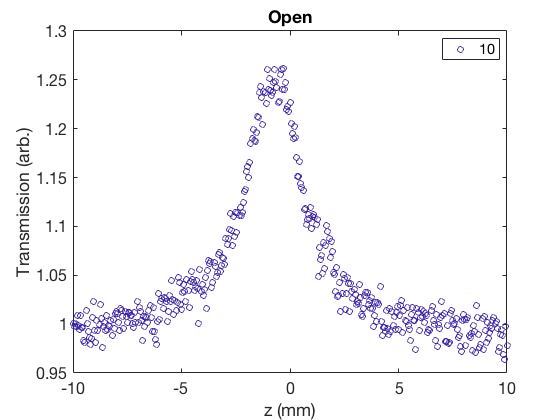}
    \includegraphics[width=0.45\textwidth]{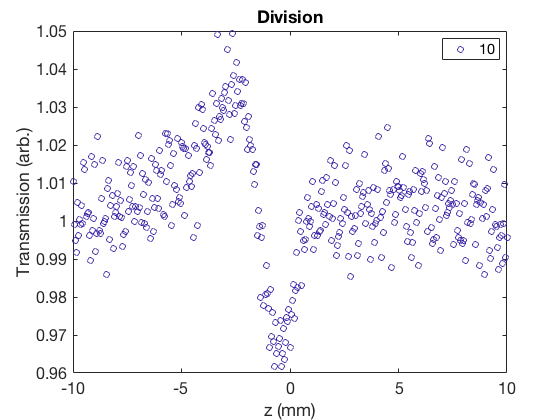}
    \caption{Experimental z-scan of 370nm thick film of pure DCDHF6V at 10µW of average incident power, pumped using wavelength of 680nm.}
    \label{fig:filmExp}
\end{figure}

\noindent \textit{Non-polaritonic thin film} \newline
Lastly, we briefly show the data corresponding to the z-scan of the pure DCDHF-6V film (380nm thick). The open and closed/open traces are shown in Fig. \ref{fig:filmExp}. Note the drastically reduced dynamic range associated with the division, and thus the significantly smaller index change when compared to the polaritonic samples. This is clear evidence of polaritonic enhancement of intensity dependent nonlinear refractive index induced by the polariton resonance.

%-----------------------------------------------------------------------------------------------------------------------------------------------------------------------------------------------------------------------------------------------------------------------------------------------------------------------------------------------

\end{document}